\documentclass[psfig,dvips,useAMS,usenatbib]{mn2e}
\usepackage{times}
\usepackage{epsfig}
\usepackage{amsmath}
\usepackage{natbib}
\usepackage{longtable}
\usepackage{psfrag}
\def\mpc{h^{-1}{\rm{Mpc}}}
\def\kms {\rm{km~s^{-1}}}
\def\apj {ApJ}
\def\apjl {ApJL}
\def\apjs {ApJS}
\def\aj {AJ}

\def\aap {A\&A}
\def\mnras {MNRAS}

\def\zspec {z_{\rm spec}}

\def\mpc{{\  h^{-1} \rm Mpc}}
\def\kpc{{\ h^{-1} \ \rm kpc}}
\def\kms{{\rm km \ s^{-1}}}

\title{Galaxy triplets in Sloan Digital Sky Survey Data Release 7: II. A connection with compact groups?}

\author[Duplancic et al.]
{Fernanda Duplancic$^{1,2}$\thanks{E-mail: fduplancic@icate-conicet.gob.ar}, Ana Laura O'Mill$^{3}$, Diego G. Lambas$^{4,2}$,  Laerte Sodr\'e Jr$^{3}$ \& \newauthor Sol Alonso$^{1,2}$ \\
    $^1$ Instituto de Ciencias Astron\'omicas, de la Tierra y del Espacio, ICATE, Casilla de Correo 49, CP 5400, San Juan, Argentina \\
  $^2$Consejo de Investigaciones Cient\'\i ficas y T\'ecnicas (CONICET),\\Avenida Rivadavia 
1917, C1033AAJ, Buenos Aires, Argentina\\
  $^3$ Departamento de Astronomia, Instituto de Astronomia, Geof\'\i sica
e Ci\^encias Atmosf\'ericas da USP,\\ Rua do Mat\~ao 1226, Cidade
Universit\'aria, 05508-090, S\~ao Paulo, Brazil.\\
$^4$ Instituto de  Astronom\'\i a Te\'orica y Experimental, IATE, Observatorio Astron\'omico, Universidad Nacional de C\'ordoba,\\ Laprida 854, X5000BGR, C\'ordoba Argentina 
}

\date{\today}

\begin{document}
\pagerange{\pageref{firstpage}--\pageref{lastpage}}

\maketitle

\label{firstpage}

\begin{abstract}

We analyse a sample of 71 triplets of luminous galaxies derived from the
work of \citet{OMill2012}. We compare the properties of triplets and their members  with those of
control samples of compact groups, the ten brightest members of rich clusters, and galaxies in pairs.

The triplets are restricted to have members with spectroscopic redshifts in the
 range $0.01\le z \le 0.14 $ and absolute r-band luminosities brighter than
 $M_r = -20.5$. For these member galaxies, we analyse the stellar mass content, the
star formation rates, the $D_n(4000)$ parameter and ($M_g-M_r$) colour index.
Since galaxies in triplets may finally merge in a single system,
we analyse different global properties of these systems. We calculate the
probability that the properties of galaxies in triplets are strongly correlated. We
also study total star formation activity and global colours, and define the triplet
compactness as a measure of the percentage of the system total area that is filled by
the light of member galaxies. We concentrate in the comparison of our results with those of compact
groups to assess how the triplets are a natural extension of these compact systems.

Our analysis suggest that triplet galaxy members behave similarly to
compact group members and galaxies in rich clusters. We also find that systems comprising three
blue, star-forming, young stellar population galaxies (blue triplets) are most probably real systems and not
a chance configuration of interloping galaxies.
The same holds for triplets composed by three red, non star-forming
galaxies, showing the correlation of galaxy properties in these systems. 
From the analysis of the triplet as a whole, we conclude
that, at a given total stellar mass content, triplets show a total star formation activity and global
colours similar to compact groups. However, blue triplets show a high total star
formation activity with a lower stellar mass content. 
From an analysis of the compactness parameter of the systems we find that
light is even more concentrated in triplets than in compact groups.

We propose that triplets composed by three luminous galaxies, should not be considered as an analogous of
galaxy pairs with a third extra member, but rather they are a natural extension of compact groups.

\end{abstract}



\section{Introduction}

It is well known that several properties of galaxies depend on environment. Pioneering work by \citet{Dressler1980} 
shows that galaxy morphology depends on local galaxy density: late type galaxies prefer low density environments, 
while dense environments tend to be populated by early type galaxies. There is also a strong 
correlation between galaxy star formation rates (SFRs) and environment: galaxies in high density environments present a decrease 
in SFRs compared to field galaxies \citep{Gomez03, Balogh04, Baldry04, mateus04}. The effect of local environment 
on galaxy colours has been studied by \citet{OMill08}. This authors find that at $z=0$ faint galaxies 
show a clear increase in the fraction of red galaxy as the local density increases. On the other hand, 
bright galaxies, present a constant red galaxy fraction. 

There are several known processes that influence galaxy properties. The hot intra-cluster gas is 
the main responsible for stripping the gas of galaxies in the core of clusters. Mechanisms such as ram-pressure stripping 
\citep{GunnGott1972} and the strangulation or starvation scenarios \citep{Larson1980} lead to a decrease 
in the star formation rate of cluster galaxies. Tidal interactions of galaxy pairs and mergers affect the morphology of galaxies and can convert
spiral galaxies into elliptical and S0s \citep{toom}. These processes can also trigger star formation, depending on the gas reservoir of the galaxies \citep{Yee-Ellingson,kenni}. In the extremely dense environments of compact groups, galaxies are separated only by a few galaxy radii from each other and with a low relative velocities, an ideal scenario for interactions and mergers \citep{Mamon1992}.

Several studies have analysed the properties of galaxies in different systems, such as pairs, compact groups, groups with four or more members, 
and clusters. Nevertheless, there are few works addressing the properties of galaxies in triple systems. \citet{HT2011} performed \textit{BVRI} 
surface photometry of a sample of 54 galaxies selected from the Catalogue of Isolated Triplets of Galaxies in the Nothern Hemisphere 
\citep{karachentseva}  and investigate the properties of 34 galaxies in 13 triplets. 
These authors found that these systems are spiral dominated and a fraction of 56\% of the triplets present morphological signatures 
associated with interactions. They also found a fraction of 35\% of bars, that can rise up to 66\% in the
late-type spirals, and a fraction of 20\% of rings, hosted preferentially in late-type components. 
Based on these results 
the authors suggest that triple systems are essentially different from Hickson Compact Groups and more representative of the field. 
However, the results of this work are based mainly in observations of a low number of systems.  
 For this reason, these authors also highlight the relevance of building a complete sample of local isolated triplet of galaxies, 
that allows a significant statistical analysis.

\citet{OMill2012} (hereafter Paper I) constructed a sample of isolated triplets of galaxies brighter than $M_r=-20.5$ and complete up to 
$z=0.4$ from the Data Release 7 of Sloan Digital Sky Survey \citep{dr7}. The aim of this paper is to analyse the properties of a sample of spectroscopic galaxies derived from this triplet catalogue and compare them with the properties of galaxies in different systems such as pairs, compact groups and clusters. 
The proximity and low relative velocities of galaxy members in triplets present an ideal scenario for 
galaxy interactions and mergers, placing these systems in a state of ongoing collapse. For these reason in this paper we will also investigate the properties of the system as a whole.

This paper is organized as follows. In Section 2 we describe the triplet and galaxy system samples used in this work. In this section we also build control samples in order to avoid biases in the comparison of properties of galaxies in different systems.
In section 3 we  analyse the specific star formation rate ($SFR/M_*$), the strength of the $4000$ \AA{} break, 
$D_n(4000)$, which is an spectral indicator of the stellar population mean age, and ($M_g-M_r$) colour index of galaxies 
in the different systems under analysis. 
We study all these properties as a function of the stellar mass content in order to avoid biases due to differences 
in this parameter. 
In section 4, we analyse de triplet as a system. We use bootstrap techniques in order to assess the probability 
that the configurations of galaxies in triple systems are correlated or, on the contrary, can be explained by a random sampling. 
We also investigate how the total stellar mass of the system affect the properties of galaxies in triplets and analyse global 
properties as total star formation rate and global ($M_g-M_r$) colour index of the system as a whole. 
We also analyse the compactness of the triplets by defining a parameter S that is a measure of the percentage of the system total 
area that is filled by the light of member galaxies. In this section we also compare the properties of triple systems with the global properties of compact groups. 
Finally in Section 6 we discuss our main results.

Throughout this paper we adopt a cosmological model characterised by
the parameters $\Omega_m=0.25$, $\Omega_{\Lambda}=0.75$ and $H_0=70~h~
{\rm km~s^{-1}~Mpc^{-1}}$.


\section{Samples}
In this work we analyse several samples of galaxy systems 
derived from the Data Release 7 of Sloan Digital Sky Survey  \citep[SDSS-DR7,][]{dr7}.
SDSS \citep{sdss} has mapped more than one-quarter of the entire sky, 
performing  photometry and spectroscopy for galaxies, quasars and 
stars. SDSS-DR7 is the seventh major data release, corresponding to the
completion of the survey SDSS-II. It comprises $11663$ sq. deg.
of imaging data, with an increment of $\sim2000$ sq. deg., over the 
previous data release, mostly in regions of low Galactic latitude.
SDSS-DR7 provides imaging data for 357 million 
distinct objects in five bands, \textit{ugriz}, as well as
spectroscopy  over $\simeq \pi$ steradians in the North Galactic 
cap and $250$ square degrees in the South Galactic cap. 
The average wavelengths corresponding to the five broad bands 
 are $3551$, $4686$, $6165$, $7481$, and $8931$ \AA{} \citep{fuku96,hogg01,smit02}. 
For details regarding the SDSS camera see \citet{gunn98}; for astrometric 
calibrations see \citet{pier03}. 
The survey has spectroscopy over 9380 sq. deg.; the spectroscopy is now 
complete over a large contiguous area of the Northern Galactic Cap, closing the gap 
that was present in previous data releases.

\subsection{Galaxy triplets sample}
In Paper I we analysed spectroscopic and photometric data extracted from SDSS-DR7 
in order to build a catalogue of isolated triplets of galaxies. 
The spectroscopic data were derived from the Main Galaxy Sample 
(MGS; \citet{mgs}). k-corrections for this sample were calculated bandshifted to $z=0.1$, using the software 
\texttt{k-correct\_v4.2} of \citet{kcorrect}. 
The photometric data were derived from the photometric catalogue constructed by 
\citet{photo}\footnote{http://www.starlight.ufsc.br/index.php?section=1}, which
contains photometric redshift  and k-correction for the photometric data of the SDSS-DR7. 
For both data sets, k-corrected absolute magnitudes were calculated from Petrosian apparent magnitudes 
converted to the AB system. 
The final catalogue comprises 1092 isolated triplets of galaxies with absolute r-band magnitude brighter than $M_r=-20.5$ in the redshift range $0.01\leq z\leq 0.4$. 
A full description of the algorithm developed to build this catalogue can be found in Paper I.

One of the aims of this paper is to analyse different properties of galaxies in triplets by including information extracted from their spectra. For this reason we have considered only triple systems composed by three spectroscopic galaxies that are close in projected separation $r_p <200\kpc$ and with a radial velocity difference $\Delta V_\mathrm{spec}<700 \kms$. Due to  completeness in the spectroscopy, we have restricted our analysis to the redshift range $0.01\leq \zspec\leq0.14$. 

In order to select isolated systems, we implement two of the isolation criteria described in Paper 1:

\begin{enumerate}
\item $N_{05}=3$
\item $N_1 \le 4$
\end{enumerate}

Here, $N_{05}$ and $N_1$ are the number of galaxies brighter than $M_r=-20.5$ within
$0.5\mpc$ and $1\mpc$, respectively, from the triplet centre considering the same restrictions on $\Delta V$ used to 
identify triplet members.

In the same way that Paper I, we calculated the distance of each triplet to the closest neighbour group/cluster, $dc$, using two catalogues: 
\citet{Zapata2009}, updated to the SDSS-DR7, for $z\le 0.1$, and for the redshift range
$0.1\leq z\leq0.14$, the GMBCG catalogue of clusters \citep{Hao2011} that includes photometric redshift information. 
For each triplet system candidate we computed the projected distance to the closest neighbour cluster considering a radial velocity restriction $\Delta V < 1000 \kms$ when spectroscopic information is available, and $\Delta V < 7000 \kms$ when  photometric redshifts are involved.

We acknowledge that the systems obtained by these conditions are consistent with a minimum distance to a rich cluster
of $dc\ge 3\mpc$, similar to the restriction  $dc\ge 5\mpc$ given in Paper I.

Under these considerations, the final sample of galaxy triplets contain 71 isolated systems with 213 spectroscopic galaxies  brighter than $M_r=-20.5$ in the redshift range $0.01\leq z\leq0.14$.

\subsection{Comparison samples}

The aim of this work is to analyse the spectroscopic properties of galaxies in triplets and compare them with the properties 
of galaxies belonging to different galaxy systems. 
Here we consider samples of galaxies in compact groups, clusters and galaxy pairs.

\subsubsection{Compact Groups}

\citet{Mcconnachie3} identified compact groups of galaxies, using the photometric data of the SDSS-DR6, through 
the implementation of the `Hickson criteria':
\begin{enumerate}
\item{$N\left(\Delta\,m = 3\right) \geq 4$;}
\item{$\theta_N \geq 3\,\theta_G$;}
\item{$\mu_e \leq 26.0$ mag\,arcsec$^{-2}$,}
\end{enumerate}
Here \noindent $N\left(\Delta\,m = 3\right)$ is the number of galaxies in the magnitude interval
$[r_1,r_1+3]$, where $r_1$ is the $r$-band magnitude of the brightest galaxy in the group. $\mu_e$ is 
the effective surface brightness of the system calculated by distributing the flux of the member 
galaxies over $\theta_G$, where $\theta_G$ is the angular diameter of the smallest 
circle encompassing the geometric centres of the galaxies in the group. 
Finally, $\theta_N$ is defined as the angular diameter of the largest
concentric circle that contains no other (external) galaxies within
this magnitude range or brighter.

\citet{Mcconnachie3} build two catalogues: Catalogue A comprising galaxies in compact groups
with petrosian $r$-band magnitude in the range $14.5\le r \le 18.0$ and Catalogue B including  galaxies in the broader magnitude range $14.5\le r \le 21.0$. Due to independent visual inspection 
by the authors of all galaxy members in Catalogue A, the contamination 
due to gross photometric errors is negligible.
This sample includes compact groups that have $\Delta v\le 1000 \kms$ where $\Delta v$ is a measure of the maximum line-of-sight velocity difference between group members for groups with more than two members with spectroscopic redshift information. 
Catalogue B includes many more groups than Catalogue A but has the disadvantage of a larger contamination due to poor photometric classification.

We will use Catalogue A as a comparison sample for this work. Nevertheless, this catalogue was constructed from the sixth release of the SDSS, therefore we have added to this catalogue spectroscopic redshift information from SDSS-DR7, and in what follows we will use systems in Catalogue A that have all their galaxies with spectroscopic measurements.

We have cross-correlated our triplet sample with this compact group catalogue in order to exclude common systems. We found 5 coincidences of our triplets with the compact groups sample. 
From these common systems, four consist on three triplet galaxies plus one galaxy with $\Delta v > 1000 \kms$. 
The remained common-system is comprised by the three triplet galaxies and one galaxy fainter than $M_r=-20.5$. We have therefore removed these 5 compact groups from the original sample.

\subsubsection{Ten first ranked cluster galaxies}
This sample was derived from the catalogue of galaxy groups of \citet{Zapata2009} updated to the SDSS-DR7. These authors implement a 
friends-of-friends algorithm with varying linking lengths 
$D_{12}=D_0R$ and $V_{12}=V_0R$, in the direction perpendicular and parallel 
to the line-of-sight, respectively, where $D_0=0.24 \mpc$ and $V_0=450 \kms$. 
The spatial scaling $R$ takes into account the variation in the space density
of galaxies in a flux-limited sample. 
From this catalogue, we selected a sample of clusters of galaxies 
considering systems with virial masses $M_{vir} >10^{14}h^{-1}M_{\odot}$ and more than 10 galaxy members. 
The galaxies belonging to these systems were ranked in $r$-band luminosity. Then we compiled a sample 
selecting the Ten First Ranked Cluster Galaxies (10FRCGs) of each system, excluding the most luminous galaxy, 
since these galaxies usually have very distinctive properties.

\subsubsection{Pairs}
The sample of galaxy pairs analysed in this work was obtained by \citet{Lambas2012}. 
These authors construct a galaxy pair catalogue selecting galaxies in the SDSS-DR7, 
with relative projected separations $r_p <25\kpc$ and relative radial velocities difference
$\Delta V<350~ \kms$. Previous studies of the team found that these limits are adequate to define
galaxy pairs with enhanced star formation activity \citep{Lambas2003, Alonso2006}.
In this work we have excluded from these sample, galaxy pairs that resides in groups, according to 
\citet{Alonso2012}. These authors analysed in detail the properties of galaxy 
interactions in high-density environments, identifying pairs that reside in
groups by cross-correlating the total galaxy pair catalogue of \citet{Lambas2012} 
with the group catalogue constructed by \citet{Zapata2009}, updated to the SDSS-DR7. 

\vspace{1cm}

In order to compare the properties of galaxies in different systems, we cross-correlate all 
the galaxy samples described above, with the derived galaxy properties from the MPA-JHU 
emission line analysis for the SDSS-DR7\footnote{Avaible at http://www.mpa-garching.mpg.de/SDSS/DR7/}. 
From this catalogue we extract several galaxy properties: as a spectral indicator of the stellar 
population mean age, we will use the strength of the $4000$ \AA{} break, $D_n(4000)$, defined as the ratio of 
the average flux densities in the narrow continuum bands 3850-3950 \AA{} and 4000-4100 \AA{} \citep{Balogh1999}. 
We also use the star formation rate (SFR) and specific star formation rates ($SFR/M_*$) according to 
\citet{Brinchmann2004} and total stellar masses ($M_*$) calculated 
from photometry \citep{Kauffmann2003}.  
 
\subsection{Control Samples}

When comparing properties of galaxies that belong to different samples it is important to consider differences in the galaxy redshift distribution. For this reason, we construct control samples using a Monte-Carlo algorithm to randomly select galaxies 
that matched the galaxy triplet redshift distribution, for the different systems samples analysed in this work. 
The distribution of $M_r$ of triplet galaxies is truncated by definition at $M_r=-20.5$ \citep{OMill2012}. In order to avoid biases due to less luminous galaxies, we restrict our analysis to galaxies brighter than  $M_r=-20.5$.

We performed a Kolmogorov-Smirnov (KS) test between the resdhift distribution of the control
samples and the redshift distribution of triplet galaxies. From this test we obtain a \texttt{p} value that represents
the probability that a value of the KS statistic will be equal or more extreme than the observed value, if the null hypothesis holds. In all cases we obtained \texttt{p}$>$0.05 for the null hypothesis that the samples were drawn from the same distribution. Table \ref{t1} summarizes the name and number of objects in the triplet galaxy sample and in the control samples analysed in this work.

\begin{table}
\center
\caption{Name and number of galaxies in the triplet and control samples.}
  \begin{tabular}{l c}
  \hline
  \hline
 name &     number of galaxies  \\
  \hline
  \hline
triplets   & 213 \\        
$cgs$   & 230\\
pairs & 472 \\ 
10FRCGs  & 2089\\
 \hline
\hline
\label{t1}
\end{tabular}
\end{table}

\begin{figure} 
\begin{picture}(250,610)
 \put(15,0){\psfig{file=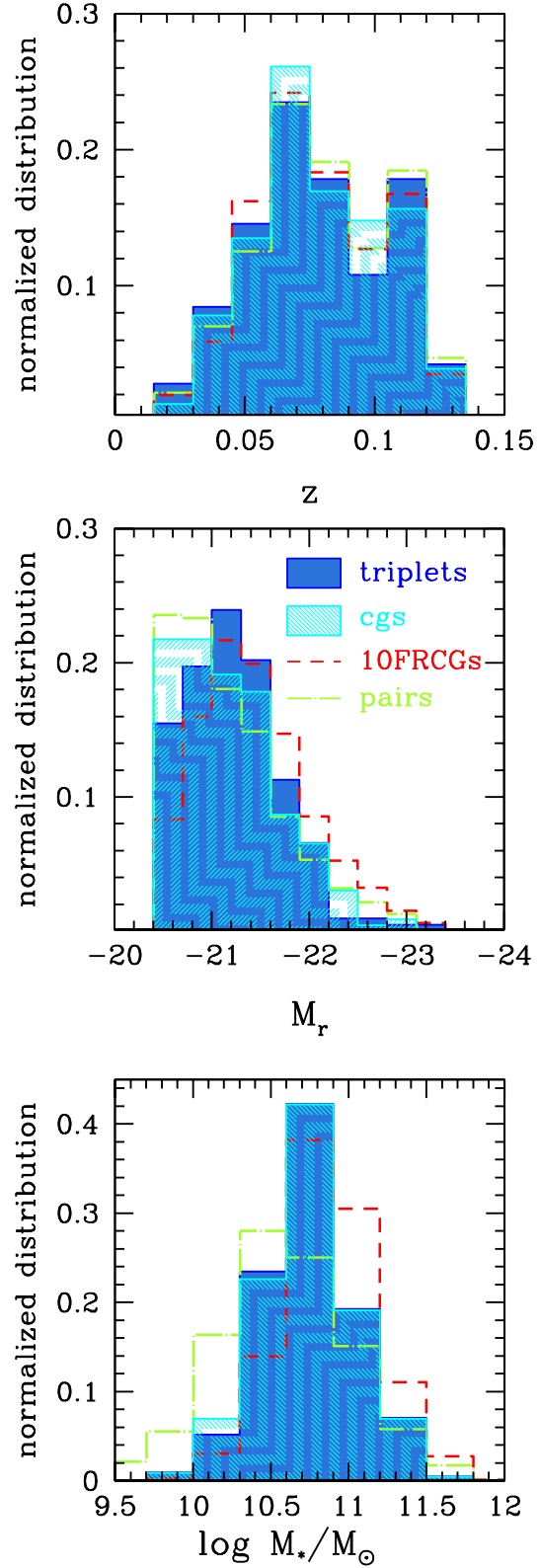, width=7cm,keepaspectratio}}
\end{picture}
    \caption{Distribution of redshift $z$ (top panel), absolute $r$-band magnitude, $M_r$ (middle panel) and log $M_*/M_{\odot}$ 
    (bottom panel) for the galaxy samples analysed in this section (key in the figure).\newline
     (A color version of this figure is available in the online journal)} 
    \label{fig1}
\end{figure}
 
Figure \ref{fig1} shows the distribution of redshift, absolute r-band magnitude ($M_r$) and log $M_*/M_{\odot}$ for the sample of triplet galaxies and for the $cgs$, pairs and 10FRCGs control samples. From this figure it can be appreciated that all the samples present a similar redshift distribution.
The restriction $M_r>-20.5$ is reflected on the log $M_*/M_{\odot}$ distribution. Although all the samples cover a similar log $M_*/M_{\odot}$ range, 
 compared to triplet galaxies, galaxies in pair presents a shift toward the less massive tail of the distribution and 10FRCGs 
show an increment in the relative number of massive galaxies with respect to the triplet sample. Nevertheless, the distribution of log $M_*/M_{\odot}$ 
for galaxies in $cgs$, is almost equal to the distribution of triplet galaxies. In order  to avoid biases due to differences in
$M_*$ we restrict our analysis to the range covered by the sample of galaxies in triplets 
($10^{10} M_\odot\leq M_* \leq 10^{11.5} M_\odot$).


\section{Analysis of the properties of galaxies in triplets}

The aim of this section is to compare the properties of individual galaxies in triplets with galaxies in compact groups, pairs and clusters. These galaxies 
reside in environments with diverse local and global densities; therefore, their main properties can be affected by different processes related to environment and evolution. 

Galaxies in clusters usually present early morphological types, low star formation rates and red colours compared with galaxies in other systems. 
The preferred scenario for their formation invokes a set of mechanisms 
that remove the gas from spiral galaxies in the cluster environment, such as 
\textit{ram-pressure stripping} of disk gas \citep{GunnGott1972} and \textit{strangulation} 
or \textit{starvation} scenarios \citep{Larson1980}.

Galaxy pair interactions play an important role in the establishment of galaxy properties. 
Different observational and theoretical analysis have shown that interactions in close pair of galaxies provide powerful mechanisms 
to trigger star formation activity \citep{Yee-Ellingson,kenni}, and the efficiency of the 
starbursts depend on the particular internal characteristics of the galaxies and of their gas reservoir \citep{toom,BH1992,BH1996,MH1996}.
In a recent work, \citet{Lambas2012}  found that, at a given total stellar mass, pairs with galaxies of similar luminosity are significantly more efficient (a factor $\approx$ 2) in forming new stars, with respect to both minor pairs (formed by two galaxies with a large relative luminosity ratio) or a control sample of non-interacting galaxies, showing that the characteristics of the interactions and the ratio of luminosity of galaxy pair members are important parameters in setting galaxy properties.

Compact group  galaxies are predominately `red and dead' \citep{Brasseur2009}. These galaxies 
reside in high local density environments, but the velocity dispersion of compact groups is 
lower than those of clusters \citep{Hickson1992}.
 This is an ideal scenario for investigating galaxy interactions 
that can affect properties such as galaxy morphology and SFR. More than 50 per cent of compact group galaxies are 
early-type \citep{Hickson1988, Palumbo1995}. Using high angular resolution observations obtained with the VLA,
\citet{VM2001} found that compact groups as a whole are HI deficient and that individual galaxies show a larger degree 
of deficiency than the groups globally (24\% of the expected HI). In most cases this could be a consequence
of efficient gas stripping from individual galaxies going into the group environment.
Although interactions between galaxies can enhance 
star formation, galaxies in compact groups present SFRs similar to those in a control sample of isolated, not strongly interacting galaxies, matched in $J$-band total galaxy luminosity \citep{Tzanavaris2010}. 
In agreement with these results, \citet{Brasseur2009} found, using simulations, 
that most compact group galaxies are red and gas-deficient, with a low specific SFR.

\begin{figure}
\centerline{\psfig{file=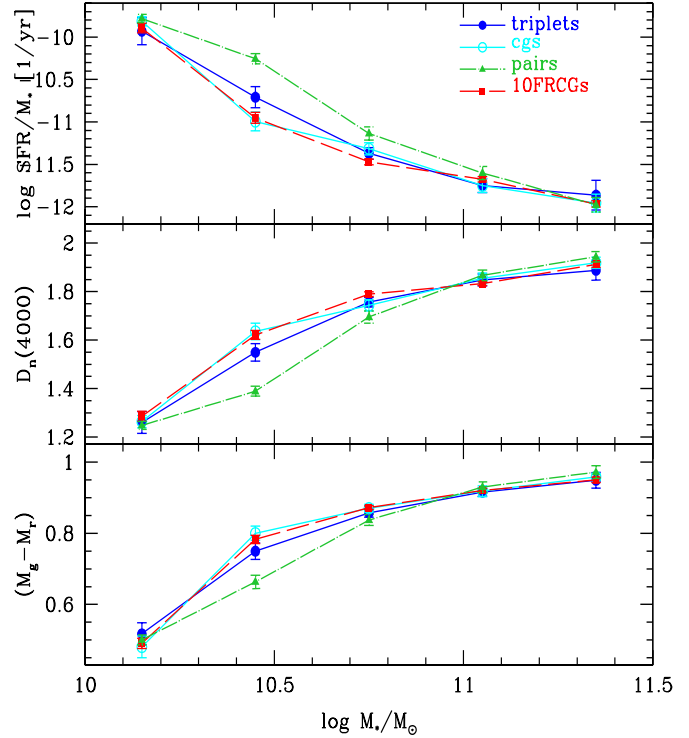,width=9cm,height=10.5cm }}
\caption{Mean values of log $SFR/M_*$ (top), $D_n(4000)$ index (middle) and $(M_g-M_r)$ colour (bottom), as a function of log $M_*/M_{\odot}$ bins, for the galaxy samples analysed in this section (key in the figure). Error bars were calculated using bootstrapping techniques.\newline
     (A color version of this figure is available in the online journal)} 
\label{f2}
\end{figure}

In order to investigate whether the galaxies in triplets 
resemble the galaxies in compact groups, or if these galaxies are similar to galaxies 
in other systems, we compute mean values of log $SFR/M_*$, $D_n(4000)$ index and $(M_g-M_r)$ colour, 
for the triplets, and for the  $cgs$, pairs and 10FRCGs control samples as a function of stellar mass (Figure \ref{f2}). 

\begin{figure*}
\begin{picture}(500,610)
\put(0,0){\psfig{file=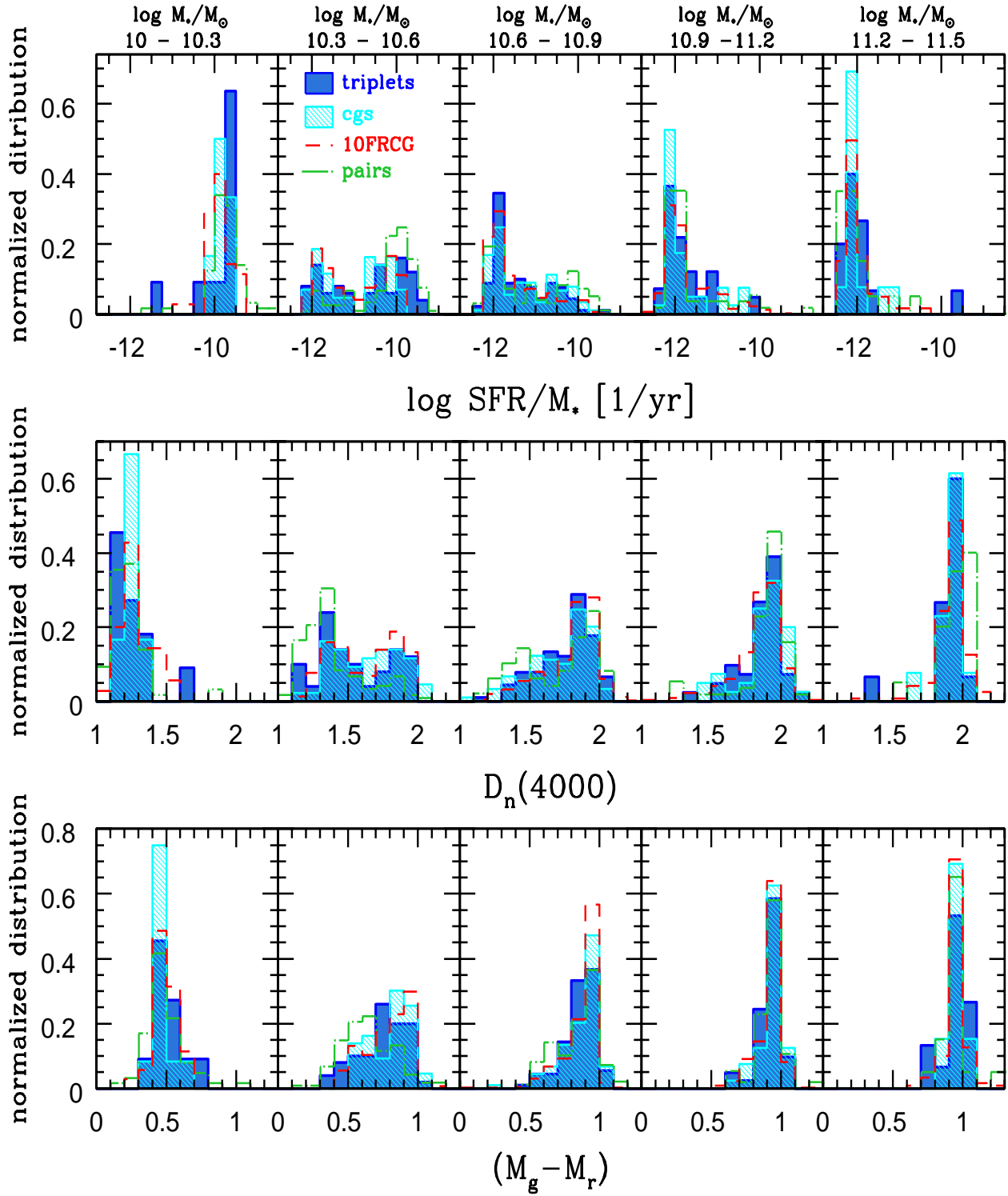 ,width=18cm,height=20cm}}
\end{picture}
\caption{Distribution of log $SFR/M_*$ (top), $D_n(4000)$ index (middle) and $(M_g-M_r)$ colour (bottom) in log $M_*/M_{\odot}$ bins, for the galaxy samples analysed in this section (key in the figure). The mean values of the distributions in each stellar mass bin are representative of the mean values of Figure \ref{f2}.\newline
     (A color version of this figure is available in the online journal) }

\label{f3}
\end{figure*}

In general, galaxies in the sample of pairs present mean values that correspond to star forming, 
blue galaxies with young stellar populations. In contrast, galaxies in triplets present a behaviour similar to 
$cgs$ and 10FRCGs: lower $SFR/M_*$, redder colours and present higher values of the $D_n(4000)$ index.

In the lower stellar mass intervals in Figure \ref{f2} it can be appreciated that galaxies step towards higher $SFR/M_*$, 
lower $D_n(4000)$ index and bluer colours. In the first stellar mass bin the mean values of the properties of galaxies in different systems are similar. For the three higher $M_*$ bins in Figure \ref{f2} there is almost 
no difference between the mean values of the properties of galaxies in the systems under analysis.

The distributions of log $SFR/M_*$, $D_n(4000)$ and $(M_g-M_r)$ colour for the galaxies in the samples analysed in different $M_*$ bins are shown in Figure \ref{f3} (log $SFR/M_*$ (top), 
$D_n(4000)$ index (middle) and $(M_g-M_r)$ colour (bottom).
We notice that the distribution of galaxies in the first $M_*$ bin is unimodal corresponding to high 
star formation (log $SFR/M_* >$-10.5), blue colours ($(M_g-M_r)<$0.75) 
and populations with low $D_n(4000)$ index ($D_n(4000)<$1.6).
For the two following stellar mass bins galaxies present a bimodal distribution  with  a 
considerable fraction of low star-forming, old stellar population, red galaxies. 
In particular, for the second $M_*$ interval, the distribution of galaxies in triplets 
has a slight tendency to resemble the distribution of galaxies in pairs, while the distribution of galaxies 
in $cgs$ is more similar to those of galaxies the in 10FRCGs sample.

When considering galaxies with stellar masses above log $M_*/M_{\odot}$=10.9, there are almost no differences between 
the distributions of the samples analysed, corresponding to a dominant population of galaxies with low SFR, high $D_n(4000)$ and red colours. This is in agreement with the results of \citet{Kauffmann2003}, who suggest that 
at stellar masses above  $3 \times 10^{10}M_\odot$ there is an increment in the fraction of old stellar 
population, bulge like, low star forming galaxies.

 From the analysis of Figures \ref{f2} and \ref{f3} we conclude that galaxies in triplets show star formation rates, colours and stellar populations that behave similarly to galaxies in compact groups and clusters. In contrast, pair galaxy members present systematically higher star formation activity indicators.


\section{Analysis of the triplet systems}

\subsection{Resampling}

\citet{Brasseur2009} suggests that the majority of galaxies in compact groups that are blue, 
gas rich and/or have high SFRs are interlopers. The results obtained in the previous section
suggest that there is a population of blue, active star forming galaxies, mainly comprised by low stellar mass objects 
in all the system samples analysed in this work. 
    
In Paper I we used a mock catalogue to analyse the completeness and contamination of
the algorithm developed for the identification of triplet of galaxies, and found a high 
level of completeness ($\sim 80 \%$) and low  contamination ($\sim 5 \%$).
Nevertheless, in order to assess the probability that blue galaxies in the triplet sample could be contaminated by interloping galaxies, 
we have attempted to determine whether the configurations of galaxies in the sample of triple systems are correlated 
or, on the contrary, can be explained by random configurations. 

For this purpose we consider blue and red triplet members ($(M_g-M_r)\leq0.75$ and $(M_g-M_r)>0.75$, respectively), star-forming and non star-forming galaxies (log $SFR/M_*\geq -10.75$ and log $SFR/M_*<-10.75$), and galaxies dominated by young and old stellar populations ($D_n(4000)\leq1.6$ and $D_n(4000)>1.6$ respectively). 

Next, we compute the number of systems that satisfy independently the following combinations of galaxy properties: 

\begin{itemize}
\item (1a) : 3 blue galaxies   
\item (1b) : 3 star-forming galaxies   
\item (1c) : 3 young stellar population galaxies
\item (2a) : 3 red galaxies   
\item (2b) : 3 non star-forming galaxies    
\item (2c) : 3 old stellar population galaxies 
\end{itemize}

We found 6 triples with combination (1a), hereafter blue triplets. These blue 
triplets also satisfy combinations (1b) and (1c). 
There are 36 triplets with combination (2a), hereafter red triplets. Out of these red 
systems, 32 belong to combination (2b) and 29 to (2c).

\begin{figure*}
\leavevmode \epsfysize=13.7cm \epsfbox{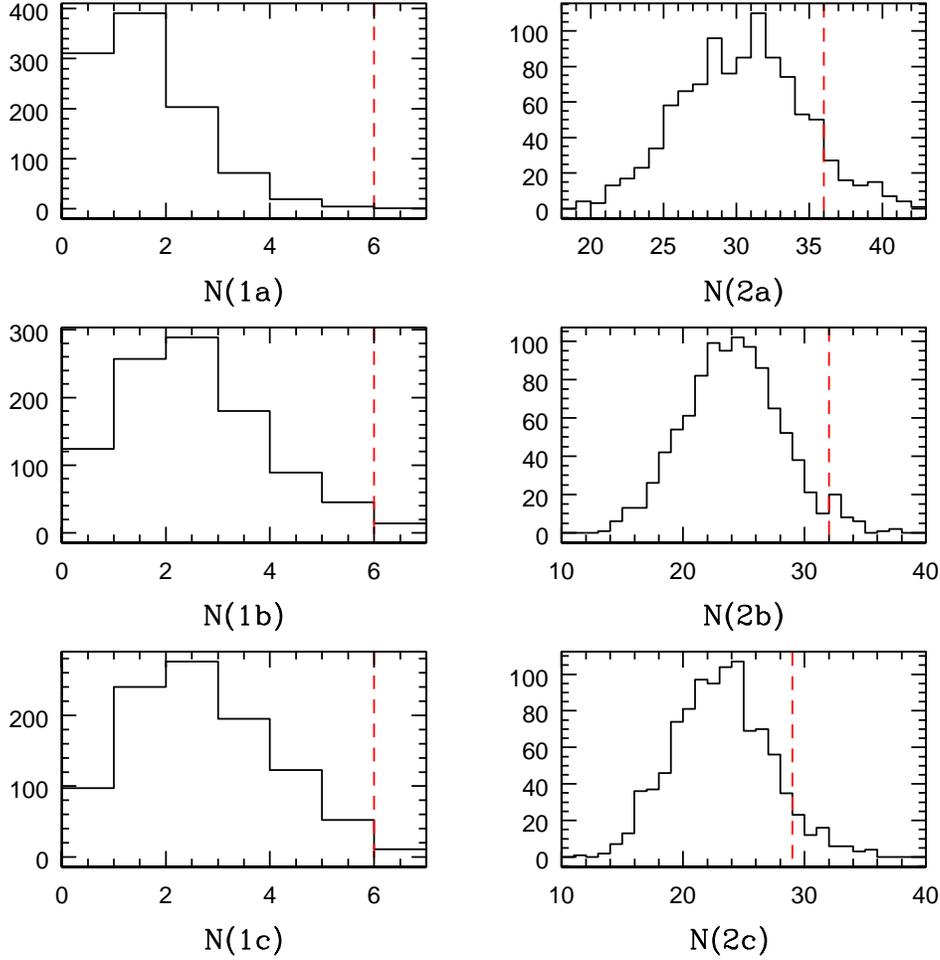}
\caption{Distribution of the number of systems that verify combinations (1a) (left upper panel), (1b) (left middle panel), (1c) (left bottom panel), (2a) (right upper panel), (2b)(right middle panel) and (2c) (right bottom panel), for 1000 triplet random samples. Vertical dashed red lines correspond to the actual values for the real triplet sample. } 
\label{f4}
\end{figure*}

In order to assess the probability that the actual number of triplets satisfying a given combination of member properties could be obtained from a random sampling, we use a bootstrap technique. Thus, we generate 1000 random triplet catalogues by a random reassignment of the galaxies from the total sample, and for each realisation we calculate the number of random triplets that satisfy the different combinations described above. This procedure allows to obtain the expected number of systems that would have these combinations by chance, and therefore assess the correlation of triplet member properties.

Figure \ref{f4} shows the distribution of the number of systems that satisfy the combinations of galaxy properties described above from the random triplet sample . 
If the combination of galaxy properties of the real triplet sample were at random, the most probable number of systems 
verifying the combinations (1a) is 1 and for (1b) and (1c) is 2. For the combination 
(2a) the mean of the distribution is 30 and for (2b) and (2c), 23.

We calculate the probability that the number of triplets satisfying each combination can be derived by a random sampling as: $p=N_{rec}/N_t$, where $N_{rec}$ is the number of times that the actual value is obtained in $N_t=1000$ trials.

We find that the probability of random occurrence of 6 systems satisfying combinations (1a) is 0.1\%; for the combination (1b) 
 1.1\%, and for (1c)  1.4\%. 
The probability of random occurrence of 36 systems satisfying combination (2a) is 2.7\%, 
that of 32 systems for combination (2b) is 2.3\% and for 29 systems 
with combination (2c) is 2\%. In table \ref{t2} we summarize these numbers.

\begin{table}
\caption{ Combination name, actual number of triples satisfying each combination and the probability $p$ 
of these combinations computed using bootstrap resampling techniques.}
  \begin{center} 
  \begin{tabular}{l c c}
  \hline
  \hline

Combination &  Number of triplets & Probability\\
 \hline
(1a)  & 6    & 0.1\% \\
(1b)  & 6    & 1.1\% \\
(1c)  & 6    & 1.4\%\\
(2a)  & 36  & 2.7\%\\
(2b)  & 32  & 2.3\%\\
(2c)  & 29 & 2\%\\
\hline
\hline
\label{t2}
\end{tabular}
\end{center}
\end{table}

The results obtained from this analysis indicate that both blue triplets and red triplets have such a 
combination of galaxies that can not be explained by a random reassignment of the galaxies. 
In particular, we conclude that systems comprising blue, star forming, young stellar population galaxies, 
have a high probability of being real systems and not a mere configuration of interloping galaxies. 
In support of this result, \citet{HT2011} found that galaxy triplets are spiral rich systems populated mostly by 
late-type spirals with an excess of $\sim 0.6$ mag in the  global blue luminosity in comparison to field galaxies.


\subsection{ Total stellar mass of the triplets and its influence on member properties}

 In the previous section we found that the galaxy configurations of triplets can not be explained by a random sampling of 
the triplet galaxies. In particular there are six blue triplets that are comprised by blue, young stellar population and 
star forming galaxies. In this section we investigate the dependence of triplet members properties on the total stellar mass of the system considering separately blue and red triplets and by comparison to the results from compact groups. 
This joint analysis of our results and those for compact groups is based on their similar stellar mass range, a fact that does not hold for pairs nor clusters.

\begin{figure}
\centerline{\psfig{file=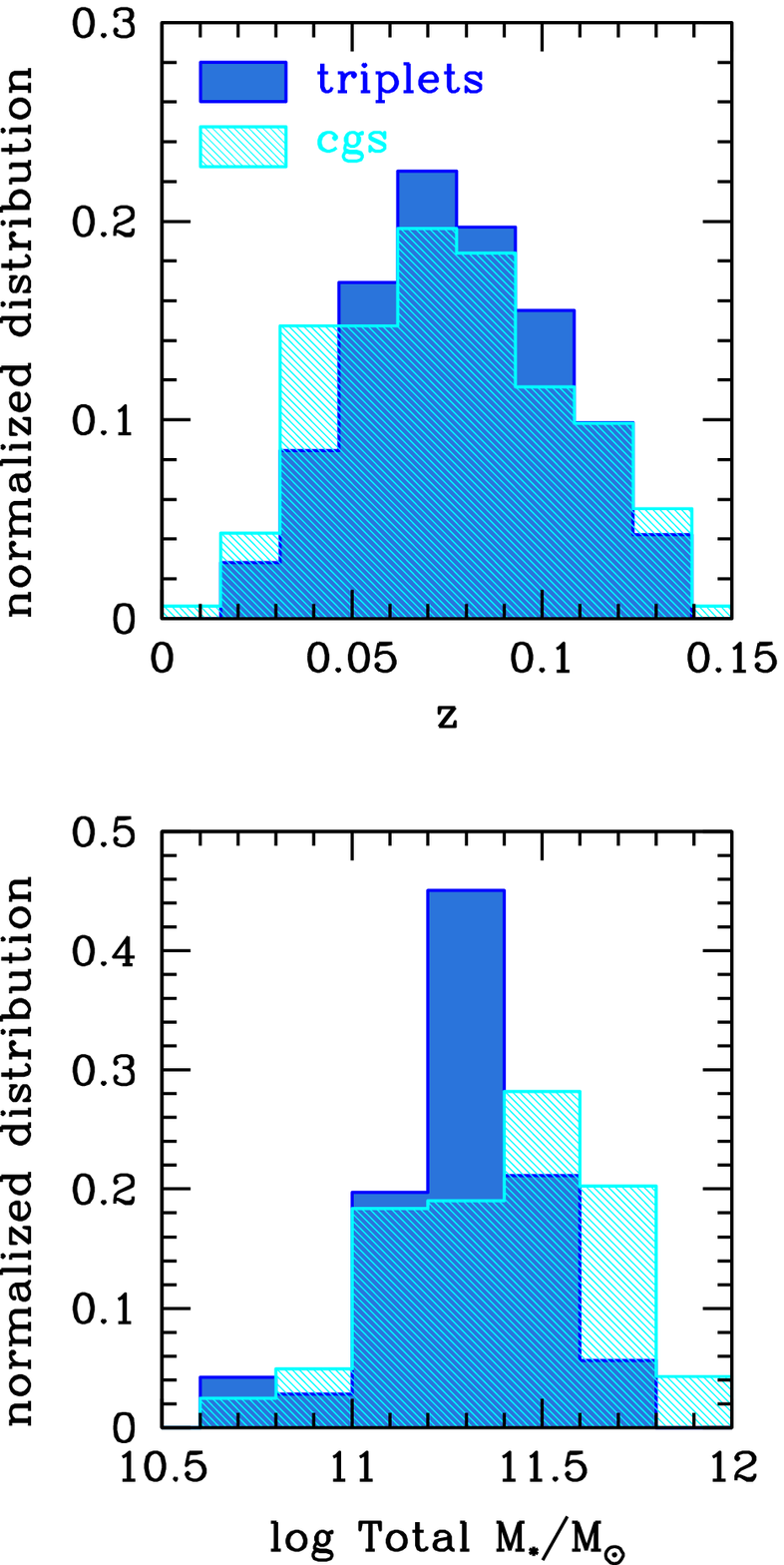,width=7.2cm,height=14.4cm }}
\caption{Distribution of redshift (top) and log Total $M_*/M_{\odot}$ (bottom) for triplets and compact groups. KS test gives a 
\texttt{p}$>0.05$ for the null hypothesis that the redshift of triplets and compact groups 
were drawn from the same distribution.\newline
     (A color version of this figure is available in the online journal)
} 
\label{f5}
\end{figure}

For the samples of triplets and compact groups described in Section 2.1 and 2.2.1, we calculate the redshift of each system as the mean redshift of their member galaxies. We restrict our analysis to the redshift range of triplets and perform a KS test to the redshift distributions of triplets and compact groups. The value obtained  for
\texttt{p} is greater than 0.05 so we can not reject the null hypothesis that the redshift of triplets and compact groups are drawn from the same distribution. The samples of triplets and compact groups used in this section comprises 71 and 163 systems, respectively and we show in Figure \ref{f5} the distributions of redshift and log Total $M_*/M_{\odot}$ for these samples. It can be appreciated that compact groups have a higher stellar mass content than triplets. This is somewhat expected since 
triplets consist of three bright galaxies while compact groups are composed by more than four members. In our study, we have considered the total stellar mass range spanned by the triplet sample (10.6$<$log Total $M_*/M_\odot <$11.8) and analyse the properties of galaxies as a function the total stellar mass content of the system. 

\begin{figure}
\begin{picture}(190,610)
\put(12,400){\psfig{file=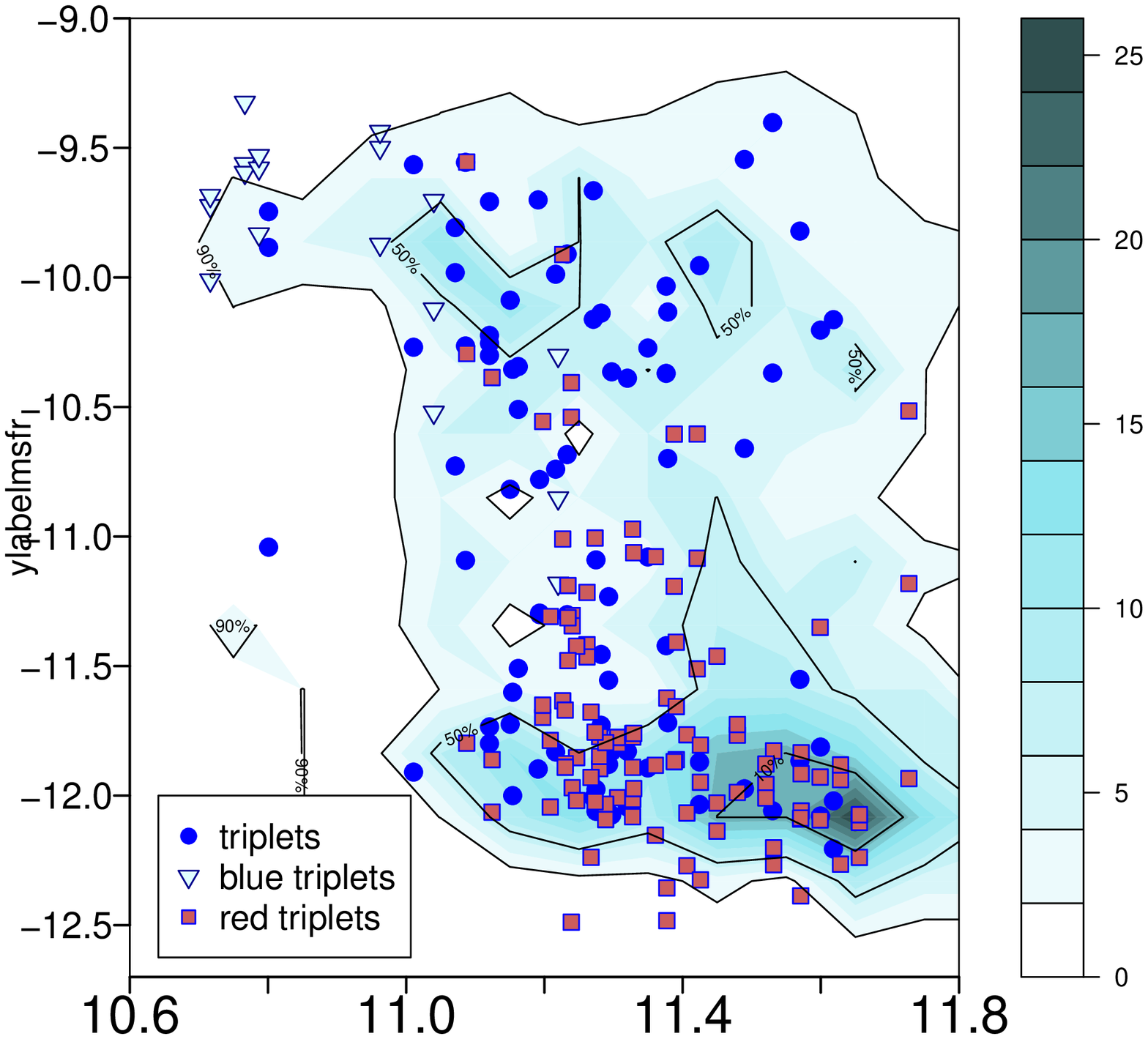 ,width=8.5cm,height=8cm}}
\put(12,200){\psfig{file=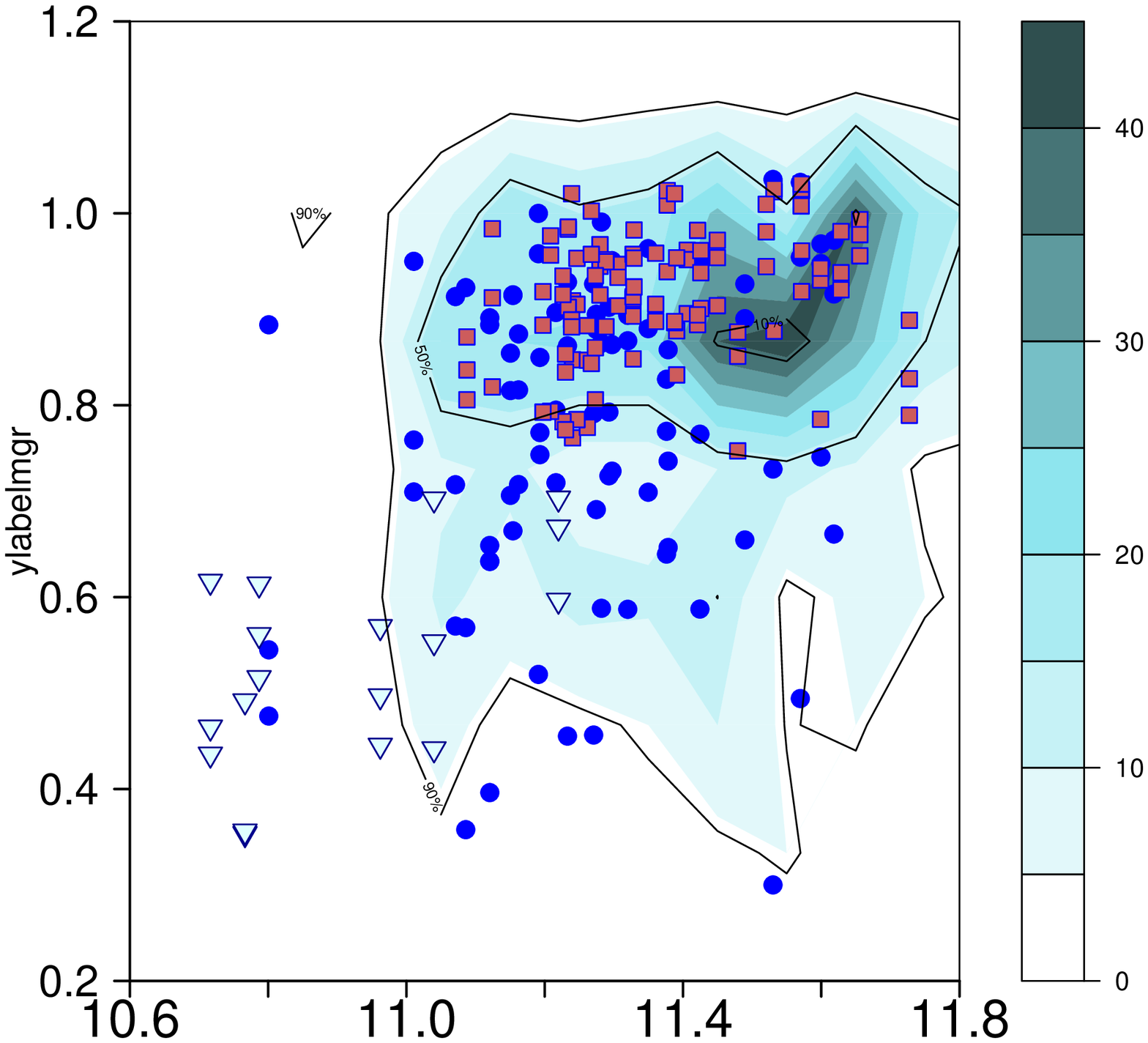,width=8.5cm,height=8cm}}
\put(12,0){\psfig{file=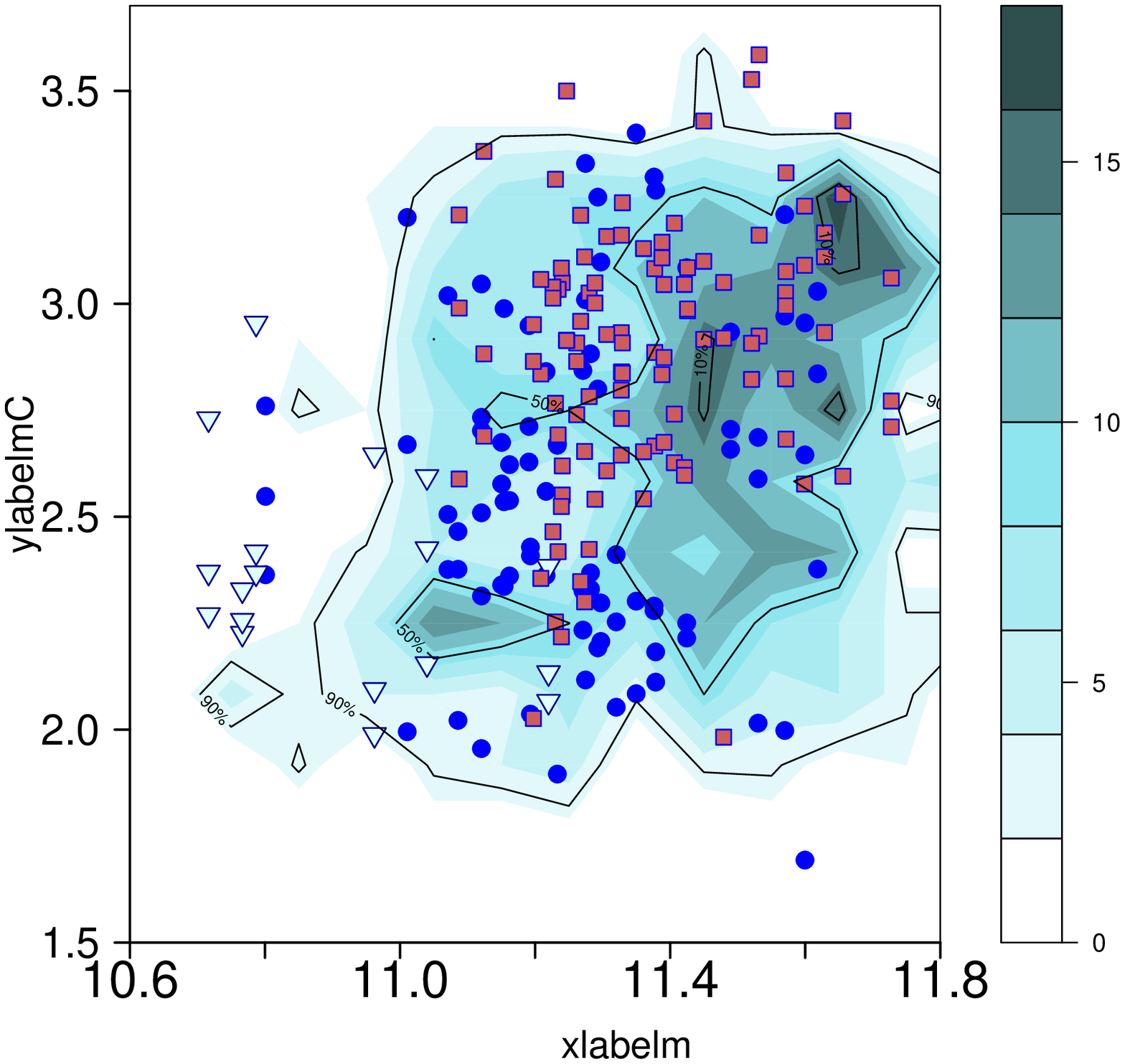,width=8.5cm,height=8cm}}
\end{picture}
\caption{$SFR/M_*$ (top), ($M_g-M_r$) colour index (middle) and concentration index $C$ (bottom) as a function of total stellar mass. 
Filled contours represent compact group galaxies, in black contours we have plotted the levels comprising the 
10\%, 50\% and 90\% of the galaxies in this sample. The points correspond to triplet galaxies where we show 
blue and red triplets as defined in Section 4 (key in the figure).\newline
     (A color version of this figure is available in the online journal)}
\label{f6}
\end{figure}

 The contours shown in Figure \ref{f6} represent the properties of compact groups members 
and the points correspond to galaxies in triplets. We have considered separately blue triplets (light-blue filled triangles), and red triplets (red filled squares), as well as triplets that do not fulfil these categories (blue dots). The top, middle and bottom panel of this figure correspond to the $SFR/M_*$, the ($M_g-M_r$) colour index and the concentration index $C$ \footnote{$C=r_{90}/r_{50}$, where $r_{90}$ and $r_{90}$ are the radii containing 90\% and 50\% of the Petrosian galaxy light}
, respectively, as a function of the total stellar mass content. 
The $C$ parameter is a suitable indicator of galaxy morphological 
type bimodality: early-type galaxies have $C>2.6$ while later type galaxies have typically $C<2.6$ \citep{Strateva2001} 

    \psfrag{xlabelm}[c][c][1.3]{log Total $M_*$/$M_\odot$}
    \psfrag{ylabelmC}[c][c][1.3]{C}
    \psfrag{ylabelmgr}[c][c][1.3]{($M_g-M_r$)}
    \psfrag{ylabelmsfr}[B][c][1.3]{log SFR/$M_*$ [1/yr]}

It can be appreciated, as expected, that the compact groups tend to have a tail towards larger total stellar masses. Nevertheless, triplets and compact groups members present a similar behaviour.
From this figure, we can also observe that blue triplets	correspond to low total stellar mass systems. Galaxies in these triplets also show a higher star formation activity and lower values of the concentration index, $C$, 
corresponding to disk objects. On the other hand, galaxies in red triplets prefer higher total stellar mass systems and tend to 
be ellipticals with a low star formation efficiency.

\subsection{Global properties of Triplets}
The dynamic of galaxy systems suggests that interactions are common in compact systems such as triplets and 
compact groups \citep{HT2011}. \citet{Diaferio94} performed dissipationless N-body simulations finding that compact 
configurations are continually replaced by new systems because within a single rich collapsing group, compact groups of 
galaxies form continuously. From the observational 
point of view, \citet{Vaisanen2008} combined near-infrared imaging and optical
spectroscopy with archival Hubble Space Telescope imaging and Spitzer imaging and spectroscopy, revealing that 
the luminous IR galaxy (LIRG) IRAS 19115-2124 is actually a triple system where the LIRG phenomenon is dominated by 
the smallest of the components. 

Since galaxies in triplets and compact groups may finally end in a single system,
in this section we analyse different global properties of triplets of galaxies by 
considering the system as a whole. We will also perform the same analysis on compact groups 
in order to compare the results obtained for these systems. For this purpose we used the samples of triplets and compact groups defined in the previous section.

With the aim to complement the previous analysis, we also explore the dependence of the total star formation rate 
and total colour as a function of the total stellar mass of the systems.To measure the level of compactness 
of the systems, we have defined a geometric criteria in order to guarantee that the systems under analysis are not 
limited by the selection criterion commonly used in determining the system compactness.
Radii of compact groups are defined by the Hickson criteria as the smallest circle that contains the 
geometric centres of compact group members. In order to define an homogeneous compactness criterion for 
triplets and compact groups we calculate the triplets minimum enclosing circle using the code of \citet{minicirc}.
Then we define the compactness $S$, as:

 $$S= \dfrac{\sum_{i=1}^{N} r_{90}^2}{r_{max}^2}$$

where $r_{90}$ is the radius enclosing $90\%$ of the Petrosian flux of the galaxy in the r-band,  
$r_{max}$ is the minimum enclosing 
circle that contains the geometric centres of the galaxies in the system and N is the total number of members of the system.
By definition, this quantity is a measure of the percentage of the system total area that is filled by 
the light of member galaxies.    

\begin{figure*}
\begin{picture}(500,610)

\put(15,390){\psfig{file=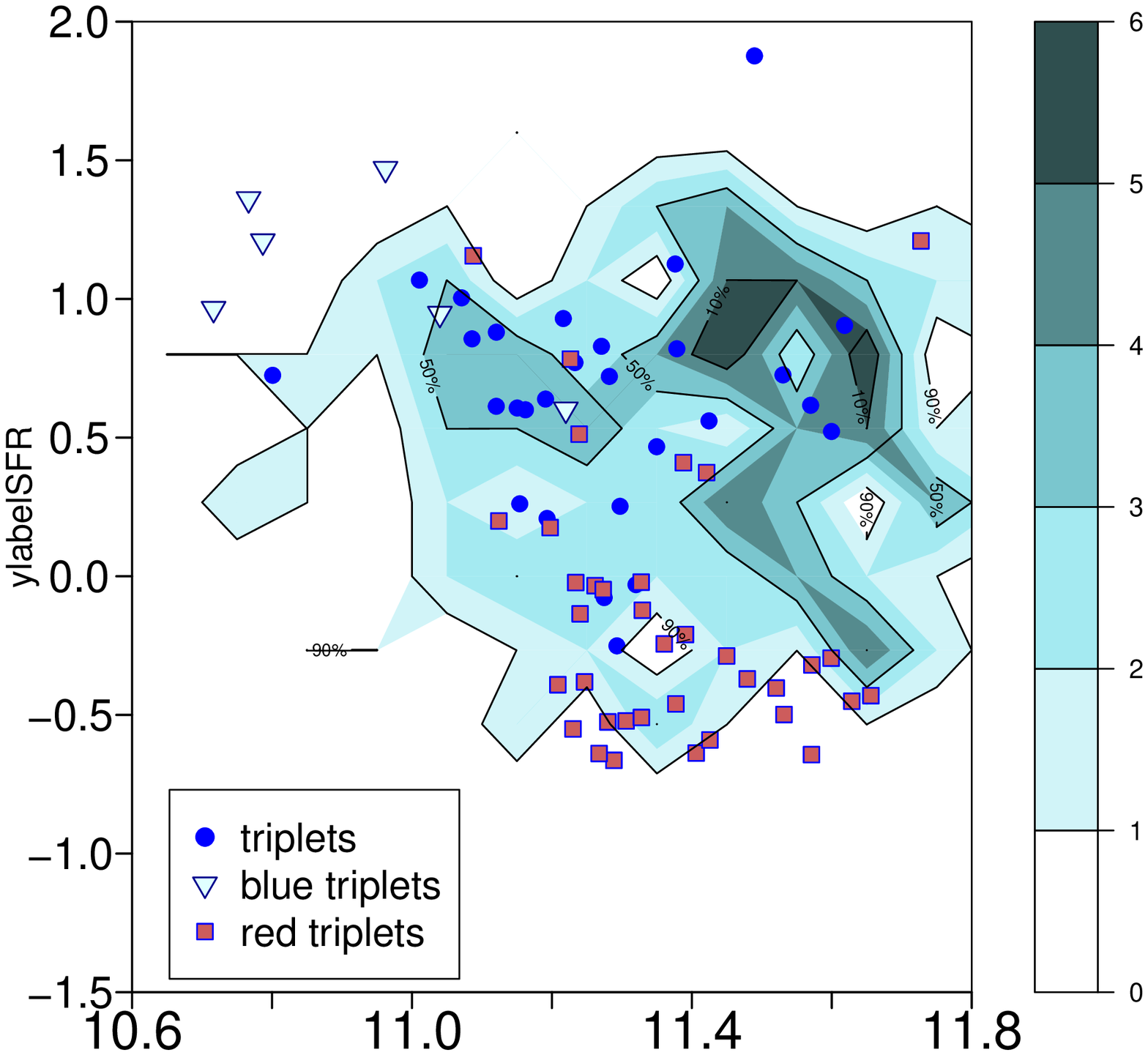 ,width=8.5cm,height=8cm}}
\put(15,195){\psfig{file=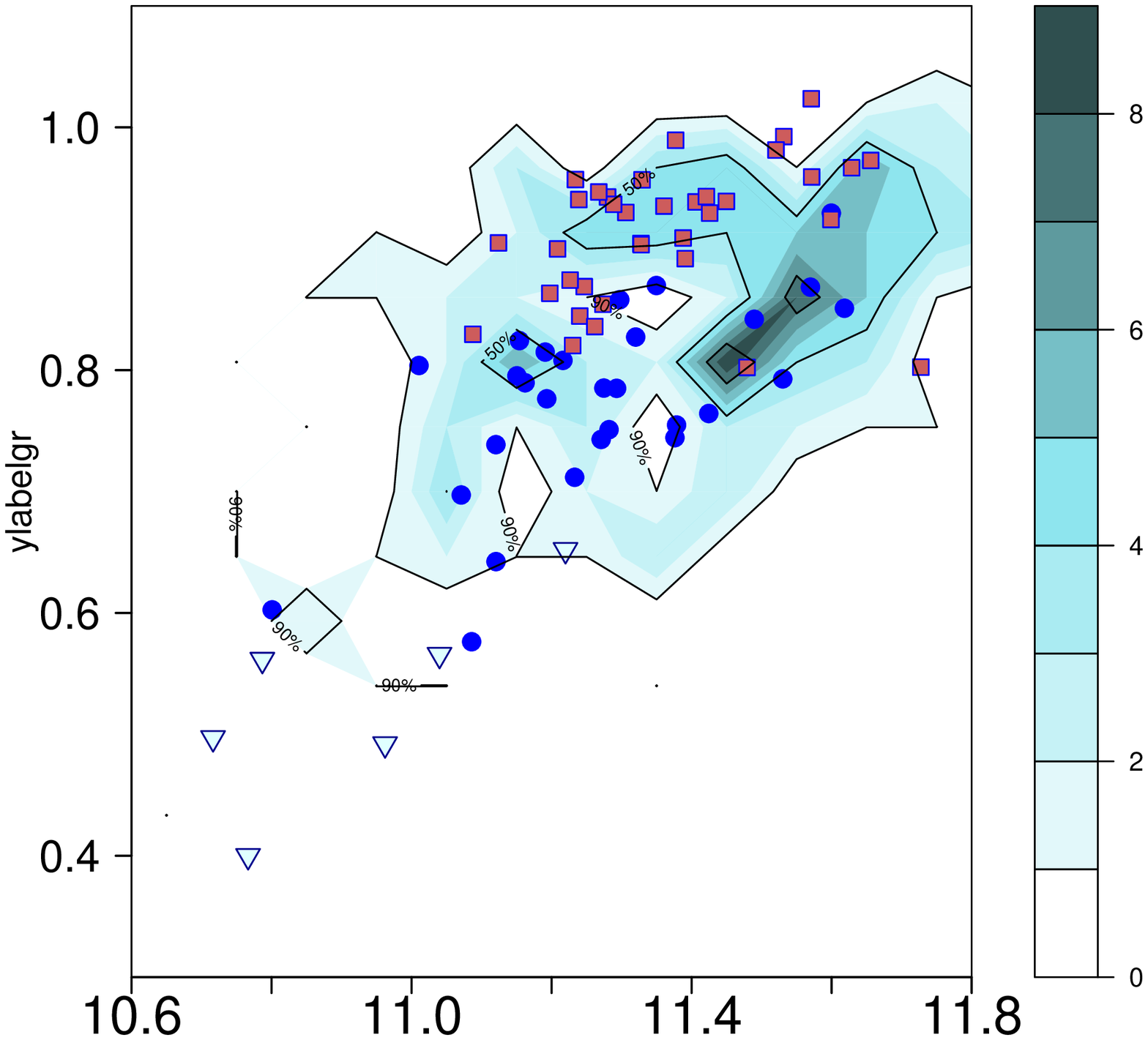,width=8.5cm,height=8cm}}
\put(15,0){\psfig{file=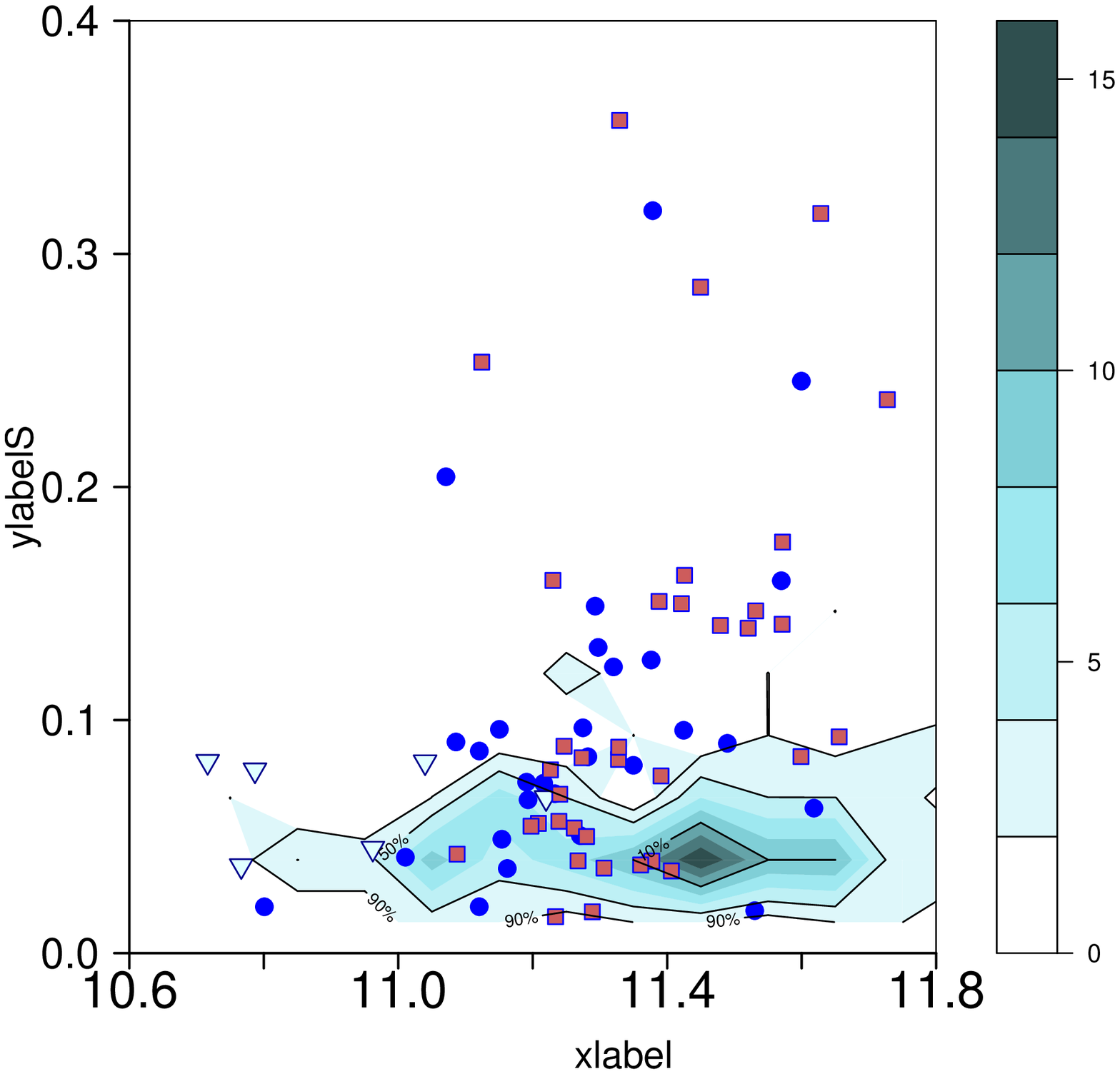,width=8.5cm,height=8cm}}
\put(270,405){\psfig{file=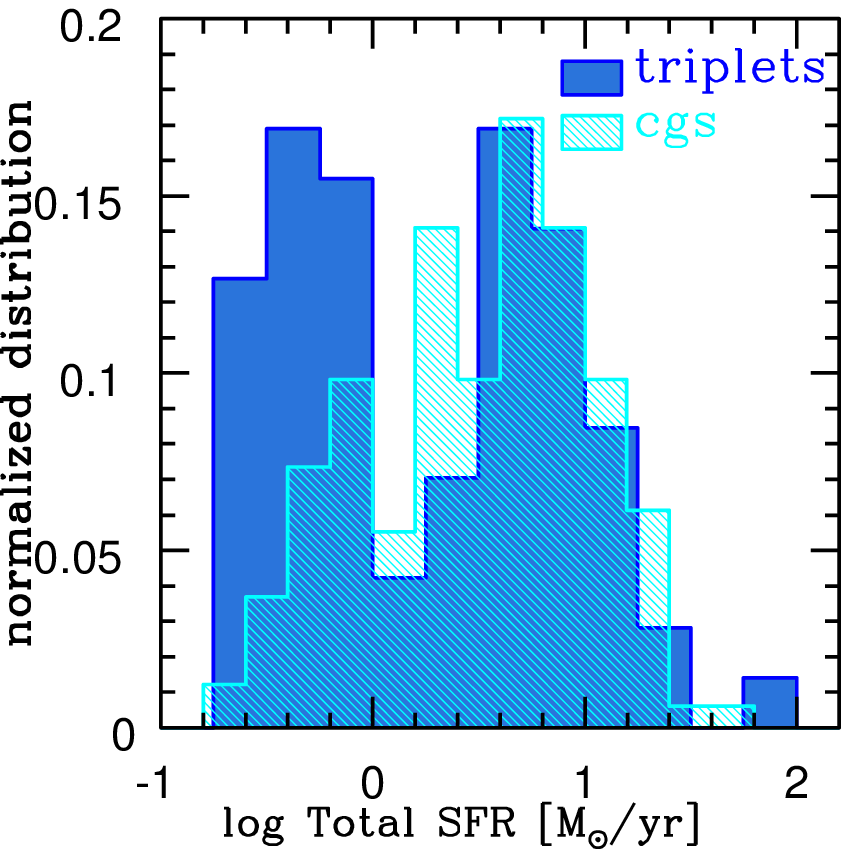 ,width=6.8cm}}
\put(270,205){\psfig{file=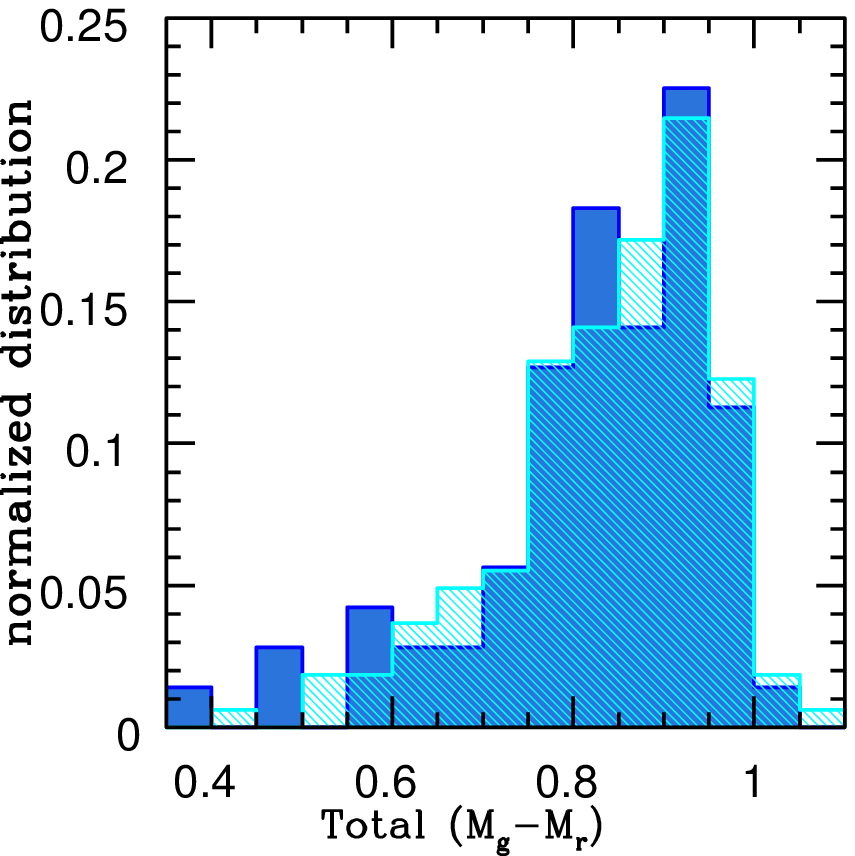,width=6.8cm}}
\put(270,10){ \psfig{file= 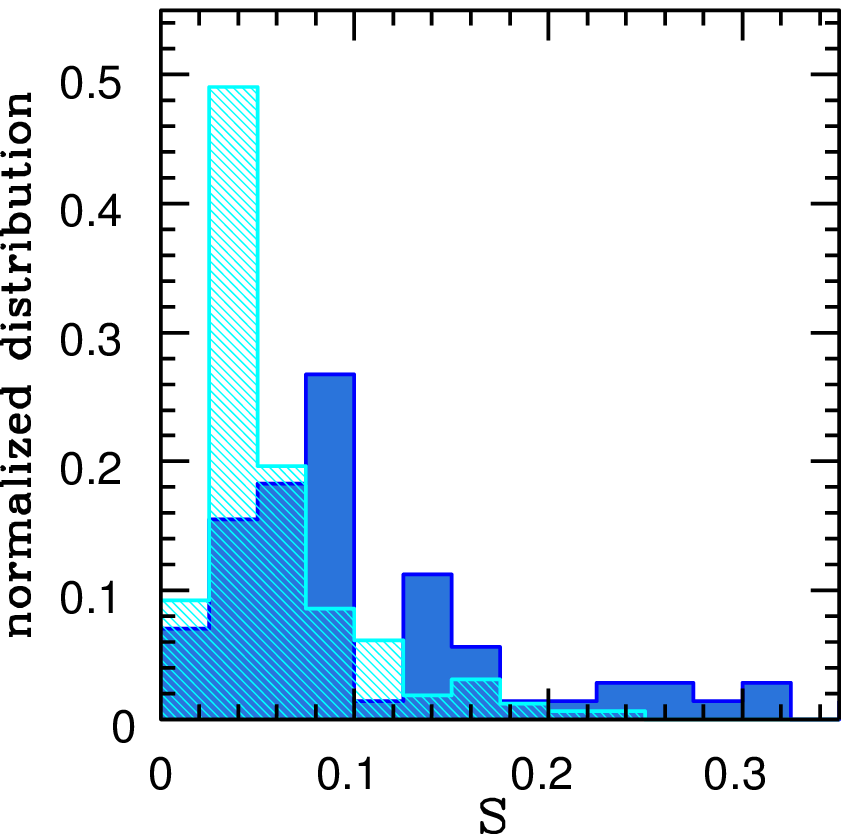,width=6.8cm}}
\end{picture}
\caption{\textit{Left}: Total SFR (top), Total ($M_g-M_r$) colour index (middle) and compactness S (bottom), as a function of 
total stellar mass. Filled contours represent compact groups, in black contours we have plotted the levels 
comprising the 10\%, 50\% and 90\% of the systems in this sample. The points indicate triplets, we have discriminated 
between blue and red triplets as defined in Section 4 (key in the figure). \newline
\textit{Right:} Total SFR (top), Total ($M_g-M_r$) colour index (middle) and compactness S (bottom) distribution for triplets and compact groups (see key in the figure).\newline
     (A color version of this figure is available in the online journal)}
\label{f7}
\end{figure*}

Left panels in Figure \ref{f7} show the properties of the systems as a whole, as a function of the total stellar mass content. 
Here again, the contours represent the compact groups, and the points correspond to the triplets.
Right panels in this figure show the distribution of the total SFR, global $(M_g-M_r)$ colour, and compactness $S$ 
for both samples.

As it was observed for individual galaxies, it can be seen that the blue triplets have a high total star formation 
activity and tend to be less massive systems. Red triplets are more massive objects and show a lower total star formation rate. 
From this figure it can also be appreciated that there is a correlation between the total star formation rate and total 
stellar mass for the triplets, while compact groups present an approximately constant trend. This is reflected in the total 
SFR distribution (right upper panel): triplets are clearly bimodal, while the compact groups distribution is more consistent with a single population.

Regarding total colour, there is a positive correlation with total stellar mass in both samples. As expected by its definition,
 blue triplets are located in the blue, low mass tail of the distribution.  
 
 In the left bottom panel of Figure \ref{f7}, we show the S parameter as a function of the total stellar mass.
As can be appreciated, the compactness S of triplets increases toward high total stellar masses, and blue triplets tend to have less compact configurations. On the other hand, compact groups show a nearly constant trend. 
Besides, as seen in the right bottom panel, it can be appreciated that the light in triplet systems is even more concentrated than in compact groups.

    \psfrag{xlabel}[c][c][1.3]{log Total $M_*$/$M_\odot$}
    \psfrag{ylabelS}[c][c][1.3]{S}
    \psfrag{ylabelSFR}[c][c][1.2]{log Total SFR [$M_\odot$/yr]}   
    \psfrag{ylabelgr}[c][c][1.2]{ Total ($M_g-M_r$)}
    
\section{Results and discussions}

We have studied 71 galaxy triplets in the redshift range $0.01\le z \le 0.14 $ with galaxy members brighter than $M_r < -20.5$. 
Our study also includes a comparative analysis of galaxies in compact groups, the ten brightest members of rich clusters, and pair galaxies.
We also analyse the stellar mass content, the star formation rates, the spectral $D_n(4000)$ index and the ($M_g-M_r$) colour index of member galaxies.

Galaxies in triplets show star formation rates, colours and stellar populations comparable to those of galaxies in compact groups and clusters. This contrasts with pair galaxy members, which have systematically higher star formation activity indicators.
Thus, we conclude that triplets cannot be considered as an extension of galaxy pairs with a third extra member.

With the aim of determining whether the configurations of galaxies in the sample of triple systems are strongly correlated or, on the contrary, can be explained by a random sampling, we generate 1000 random triplet catalogues by a random reassignment of the galaxies from the real sample.
From this analysis we conclude that systems comprising three blue, star-forming, young stellar 
population galaxies (blue triplets) are most probably real systems and not a chance configuration of interloping galaxies. 
The same holds for triplets composed by three red, non star-forming galaxies showing the correlation of galaxy properties in these systems.

Since triplets and compact group members may finally merge into a single system we have computed global parameters to provide a fair comparison of these systems as a whole.
We have analysed different properties, such as total star formation rates and total global colours, finding that, in general, triplets and compact groups exhibit a similar behaviour.
Nevertheless we find that blue triplets, located in the less massive tail of the total stellar mass distribution, show an efficient total star formation activity with respect to compact groups which present low efficiency in forming new star generations, in the same total stellar mass range.
In contrast, both, high stellar mass red triplets and compact  groups in the same stellar mass region, show low total star formation activity. 

In order to provide a compactness parameter of the triplets that can be suitably compared to the other systems, we have defined a parameter S as the sum of the areas comprising 90$\%$ of the light of galaxies divided into the system minimal enclosing circle area. This quantity is a measure of the percentage of the system total area that is filled by 
the light of member galaxies.
The distribution of this compactness parameter shows, surprisingly, that light is even more concentrated in triplets than in compact groups. 

The results obtained in this work suggest that triplets composed by three luminous galaxies, have member with properties more similar to compact group members than to galaxies in pairs, and that the behaviour of triplets as a whole is similar to compact groups. Based on the results presented in this work, we argue that triplets are a natural extension of compact groups to systems with lower number of galaxies.


\section{Acknowledgements}

We thank the referee, Nelson Padilla, for providing us with helpful comments that improved this work.
We thank the authors Zapata, Perez, Padilla \& Tissera for granting us access to their
group catalogue. This work was supported in part by the Consejo Nacional de 
Investigaciones Cient\'ificas y T\'ecnicas de la Rep\'ublica Argentina 
(CONICET), Secretar\'\i a de Ciencia y Tecnolog\'\i a de la Universidad 
 de C\'ordoba. Laerte Sodr\'e Jr and Ana Laura O'Mill were supported by the Brazilian agencies FAPESP and CNPq. 
Funding for the SDSS and SDSS-II has been provided by the Alfred P. Sloan Foundation, 
the Participating Institutions, the National Science Foundation, the U.S. 
Department of Energy, the National Aeronautics and Space Administration, 
the Japanese Monbukagakusho, the Max Planck Society, and the Higher Education 
Funding Council for England. The SDSS Web Site is http://www.sdss.org/. 
The SDSS is managed by the Astrophysical Research Consortium for 
the Participating Institutions. The Participating Institutions are 
the American Museum of Natural History, Astrophysical Institute 
Potsdam, University of Basel, University of Cambridge, Case 
Western Reserve University, University of Chicago, Drexel 
University, Fermilab, the Institute for Advanced Study, the Japan 
Participation Group, Johns Hopkins University, the Joint Institute 
for Nuclear Astrophysics, the Kavli Institute for Particle 
Astrophysics and Cosmology, the Korean Scientist Group, the 
Chinese Academy of Sciences (LAMOST), Los Alamos National 
Laboratory, the Max-Planck-Institute for Astronomy (MPIA), the 
Max-Planck-Institute for Astrophysics (MPA), New Mexico State 
University, Ohio State University, University of Pittsburgh, 
University of Portsmouth, Princeton University, the United States 
Naval Observatory, and the University of Washington.


\label{lastpage}


\end{document}